\documentclass[12pt]{article}
\usepackage{graphicx}
\usepackage{lscape}
\usepackage{citesort}
\usepackage{amssymb}
\usepackage{appendix}
\usepackage{multirow}

\begin{document}
\def\qq{\langle \bar q q \rangle}
\def\uu{\langle \bar u u \rangle}
\def\dd{\langle \bar d d \rangle}
\def\sp{\langle \bar s s \rangle}
\def\GG{\langle g_s^2 G^2 \rangle}
\def\Tr{\mbox{Tr}}
\def\figt#1#2#3{
        \begin{figure}
        $\left. \right.$
        \vspace*{-2cm}
        \begin{center}
        \includegraphics[width=10cm]{#1}
        \end{center}
        \vspace*{-0.2cm}
        \caption{#3}
        \label{#2}
        \end{figure}
    }

\def\figb#1#2#3{
        \begin{figure}
        $\left. \right.$
        \vspace*{-1cm}
        \begin{center}
        \includegraphics[width=10cm]{#1}
        \end{center}
        \vspace*{-0.2cm}
        \caption{#3}
        \label{#2}
        \end{figure}
                }

\def\ds{\displaystyle}
\def\beq{\begin{equation}}
\def\eeq{\end{equation}}
\def\bea{\begin{eqnarray}}
\def\eea{\end{eqnarray}}
\def\beeq{\begin{eqnarray}}
\def\eeeq{\end{eqnarray}}
\def\ve{\vert}
\def\vel{\left|}
\def\ver{\right|}
\def\nnb{\nonumber}
\def\ga{\left(}
\def\dr{\right)}
\def\aga{\left\{}
\def\adr{\right\}}
\def\lla{\left<}
\def\rra{\right>}
\def\rar{\rightarrow}
\def\lrar{\leftrightarrow}
\def\nnb{\nonumber}
\def\la{\langle}
\def\ra{\rangle}
\def\ba{\begin{array}}
\def\ea{\end{array}}
\def\tr{\mbox{Tr}}
\def\ssp{{\Sigma^{*+}}}
\def\sso{{\Sigma^{*0}}}
\def\ssm{{\Sigma^{*-}}}
\def\xis0{{\Xi^{*0}}}
\def\xism{{\Xi^{*-}}}
\def\qs{\la \bar s s \ra}
\def\qu{\la \bar u u \ra}
\def\qd{\la \bar d d \ra}
\def\qq{\la \bar q q \ra}
\def\gGgG{\la g^2 G^2 \ra}
\def\q{\gamma_5 \not\!q}
\def\x{\gamma_5 \not\!x}
\def\g5{\gamma_5}
\def\sb{S_Q^{cf}}
\def\sd{S_d^{be}}
\def\su{S_u^{ad}}
\def\sbp{{S}_Q^{'cf}}
\def\sdp{{S}_d^{'be}}
\def\sup{{S}_u^{'ad}}
\def\ssp{{S}_s^{'??}}

\def\sig{\sigma_{\mu \nu} \gamma_5 p^\mu q^\nu}
\def\fo{f_0(\frac{s_0}{M^2})}
\def\ffi{f_1(\frac{s_0}{M^2})}
\def\fii{f_2(\frac{s_0}{M^2})}
\def\O{{\cal O}}
\def\sl{{\Sigma^0 \Lambda}}
\def\es{\!\!\! &=& \!\!\!}
\def\ap{\!\!\! &\approx& \!\!\!}
\def\md{\!\!\!\! &\mid& \!\!\!\!}
\def\ar{&+& \!\!\!}
\def\ek{&-& \!\!\!}
\def\kek{\!\!\!&-& \!\!\!}
\def\cp{&\times& \!\!\!}
\def\se{\!\!\! &\simeq& \!\!\!}
\def\eqv{&\equiv& \!\!\!}
\def\kpm{&\pm& \!\!\!}
\def\kmp{&\mp& \!\!\!}
\def\mcdot{\!\cdot\!}
\def\erar{&\rightarrow&}
\def\olra{\stackrel{\leftrightarrow}}
\def\ola{\stackrel{\leftarrow}}
\def\ora{\stackrel{\rightarrow}}

\def\simlt{\stackrel{<}{{}_\sim}}
\def\simgt{\stackrel{>}{{}_\sim}}


\title{
         {\Large
                 {\bf
                      Thermal properties  of light tensor mesons via  QCD sum rules
                 }
         }
      }

\author{ K. Azizi$^{\dag1}$, A. T\"urkan$^{*2}$, E. Veli Veliev$^{*3}$, H. Sundu$^{*4}$ \\
$^{\dag}$Department of Physics,  Faculty of Arts and Sciences,
Do\u gu\c s University
  \\Ac{\i}badem-Kad{\i}k\"oy, 34722 Istanbul, Turkey\\
 $^{*}$Department of Physics, Kocaeli University, 41380 Izmit,
Turkey\\
  $^1$e-mail:kazizi@dogus.edu.tr\\
$^2$email:arzu.turkan1@kocaeli.edu.tr\\
$^3$e-mail:elsen@kocaeli.edu.tr\\
$^4$email:hayriye.sundu@kocaeli.edu.tr}
\date{}

\begin{titlepage}
\maketitle
\thispagestyle{empty}

\begin{abstract}
The thermal properties of $f_{2}(1270)$, $a_{2}(1320)$ and $K_2^*(1430)$ light tensor mesons are investigated in the framework of QCD sum rules at finite temperature.
In particular, the masses and decay constants of the light tensor
mesons are calculated taking into account the new operators appearing at finite
temperature. The numerical results show that at the point   which the temperature-dependent continuum threshold vanishes, the decay
constants  decrease with amount of $(70-85)\%$ compared to their vacuum values, while the
masses diminish about $(60-72)\%$ depending on the kinds of the mesons under consideration. The results obtained at zero temperature are in 
good consistency with the experimental data as well as existing  theoretical predictions.
\end{abstract}

~~~PACS number(s): 11.10.Wx, 11.55.Hx, 14.40.Be, 14.40.Df
\end{titlepage}

\section{Introduction}
The study of  strong interaction at low  energies is one of
the most important problems of the high energy physics.  This can play a crucial role to explore the structure of mesons,
baryons  and vacuum properties  of strong interaction.  The tensor
particles can provide a different perspective for understanding the
low energy QCD dynamics.  In recent decades  there have been  made great
efforts both experimentally
 and
theoretically 
 to
investigate the tensor particles in order to understand their nature
and internal structure.

The investigation of hadronic properties at finite baryon density and
temperature in QCD also plays an essential role for interpretation of
the results of heavy-ion collision experiments and obtaining the QCD phase
diagram. The Compressed Baryonic Matter (CBM) experiment of the
FAIR project at GSI is important for understanding the way of Chiral
symmetry realization  in the low energy region and,
consequently, the confinement of QCD. According to thermal QCD, 
the hadronic matter undergoes to quark gluon-plasma phase  at
a critical temperature. These kind of phase may exist
in the neutron stars and early universe. Hence, calculation of the parameters of hadrons via thermal QCD may provide us with useful information on these subjects.

 The restoration
of Chiral symmetry at high temperature requires the
medium modifications of hadronic parameters \cite{G.Brown}. There are many non-perturbative approaches to hadron physics. 
 The QCD sum rule method \cite{Shifman1,Shifman2} is one of the most attractive and applicable tools in this respect.  
 In this approach, hadrons are represented by their interpolating quark currents  and the correlation function of these
 currents is calculated using the operator product expansion (OPE). The thermal version of this approach is 
based on some basic assumptions so that the Wilson expansion and the quark-hadron duality approximation
 remain valid, but the vacuum condensates are replaced by their thermal expectation values  \cite{Bochkarev}.
 At finite temperature, the Lorentz invariance is broken  and due to the residual $O(3)$ symmetry, some new operators appear in the Wilson
 expansion \cite{Hatsuda,Mallik2,Shuryak}. These operators are expressed in terms of  the four-vector velocity of the medium and the
 energy momentum tensor. There are
numerous works in the literature on the medium modifications of
parameters of (psedo)scalar  and  (axial)vector mesons using different theoretical approaches e.g. Chiral model
\cite{Mallik}, coupled channel approach \cite{Waas,Tolos} and QCD sum rules \cite{Hatsuda,Mallik2,Veliev,Veliev3,Velievuc,Klingl,Loewe,Loewe1,arzu}. Recently, we applied this
method to calculate some hadronic parameters related to the charmed $D_2^*(2460)$ and charmed-strange $D_{s2}^*(2573)$   tensor  \cite{arzu2} mesons.

In the present work we investigate the properties of light $a_{2}(1320)$, $f_{2}(1270)$ and $K_2^*(1430)$ tensor mesons in the framework of QCD sum rules at finite temperature. We also compare 
the results obtained at zero temperature with the predictions of  some previous studies on the parameters of the same mesons in vacuum  \cite{Aliev,Ebert,Aliev2}.

The present article is organized as follows. In next section, considering the new operators raised at finite temperature,  we
evaluate the corresponding thermal correlation function to obtain the QCD sum rules for the parameters of the mesons under consideration. Last section is devoted to the numerical analysis of
the sum rules obtained  as well as  investigation of the sensitivity of the masses
and decay constants of the light tensor mesons on temperature. 

\section{Thermal QCD Sum Rules for Masses and Decay Constants of Light Tensor Mesons}
In this section we present the basics of the thermal QCD sum rules
and apply this method to some light tensor mesons like
$f_{2}(1270)$, $a_{2}(1320)$ and $K_2^*(1430)$  to compute their
mass and decay constant. The starting point is to consider the following  thermal correlation function:
\begin{eqnarray}\label{correl.func.101}
\Pi _{\mu\nu,\alpha\beta}(q,T)=i \int d^{4}x e^{iq\cdot (x-y)}
Tr\left\{\rho {\cal
T}\left[J_{\mu\nu}(x)\overline{J}_{\alpha\beta}(y)\right]\right\},
\end{eqnarray}
where $\rho=e^{-\beta H}/ Tr(e^{-\beta H})$ is the thermal density
matrix of QCD, $\beta=1/T$ with $T$ being temperature, $H$ is the QCD Hamiltonian, ${\cal T}$ indicates the time
ordered product and $J_{\mu\nu}$ is the interpolating current of
 tensor mesons.  The interpolating fields for these mesons can be written as
\begin{eqnarray}\label{tensorcurrent1}
J _{\mu\nu}^{K_2^*}(x)=\frac{i}{2}\left[\bar s(x) \gamma_{\mu}
\olra{\cal D}_{\nu}(x) d(x)+\bar s(x) \gamma_{\nu}  \olra{\cal
D}_{\mu}(x) d(x)\right],
\end{eqnarray}
\begin{eqnarray}\label{tensorcurrent2}
J _{\mu\nu}^{f_2}(x)&=&\frac{i}{2\sqrt{2}}\Big[\bar u(x)
\gamma_{\mu} \olra{\cal D}_{\nu}(x) u(x)+\bar u(x) \gamma_{\nu}
\olra{\cal D}_{\mu}(x) u(x)
\nonumber\\
&+& \bar d(x)\gamma_{\mu} \olra{\cal D}_{\nu}(x) d(x)+ \bar
d(x) \gamma_{\nu} \olra{\cal D}_{\mu}(x) d(x)\Big],
\end{eqnarray}
and
\begin{eqnarray}\label{tensorcurrent2}
J _{\mu\nu}^{a_2}(x)&=&\frac{i}{2\sqrt{2}}\Big[\bar u(x)
\gamma_{\mu} \olra{\cal D}_{\nu}(x) u(x)+\bar u(x) \gamma_{\nu}
\olra{\cal D}_{\mu}(x) u(x)
\nonumber\\
&-& \bar d(x)\gamma_{\mu} \olra{\cal D}_{\nu}(x) d(x)- \bar
d(x) \gamma_{\nu} \olra{\cal D}_{\mu}(x) d(x)\Big],
\end{eqnarray}
where 
$ \olra{\cal D}_{\mu}(x)$ denotes the derivative with respect to
four-$x$ simultaneously acting on left and right. It is
given as
\begin{eqnarray}\label{derivative}
\olra{\cal D}_{\mu}(x)=\frac{1}{2}\left[\ora{\cal D}_{\mu}(x)-
\ola{\cal D}_{\mu}(x)\right],
\end{eqnarray}
where
\begin{eqnarray}\label{derivative2}
\overrightarrow{{\cal
D}}_{\mu}(x)=\overrightarrow{\partial}_{\mu}(x)-i
\frac{g}{2}\lambda^aA^a_\mu(x),\nonumber\\
\overleftarrow{{\cal
D}}_{\mu}(x)=\overleftarrow{\partial}_{\mu}(x)+
i\frac{g}{2}\lambda^aA^a_\mu(x),
\end{eqnarray}
with  $\lambda^a (a=1,8)$ and  $A^a_\mu(x)$ are being the Gell-Mann matrices
and external  gluon fields, respectively. The currents
contain derivatives with respect to the space-time, hence we
consider the two currents at points $x$ and $y$  in
Eq. (\ref{correl.func.101}), but for simplicity, we will set $y=0$ after applying derivative with respect to $y$.

It is well known that in  thermal QCD sum rule approach, the
thermal correlation function can be calculated in two different
ways. Firstly, it is calculated in terms of  hadronic
parameters such as masses and decay constants. Secondly, it is
calculated in terms of the QCD parameters such as quark masses,
quark condensates and quark-gluon coupling constants. The coefficients
of  sufficient structures  from both representations of the same correlation
function are then equated to find the sum rules for the physical quantities under consideration. We apply
Borel transformation and continuum subtraction to both sides of the sum rules in order to further suppress the
contributions of the higher states and continuum.

Let us focus on the calculation of the hadronic side of the
correlation function. For this aim we insert a
complete set of intermediate physical  state having the same
quantum numbers as the interpolating current into Eq.
(\ref{correl.func.101}). After
performing integral over four-$x$ and setting $y=0$, we get

\begin{eqnarray}\label{phen1}
\Pi _{\mu\nu,\alpha\beta}(q,T)=\frac{{\langle}0\mid  J _{\mu\nu}(0)
\mid K_2^*(f_2)(a_2)\rangle \langle K_2^*(f_2)(a_2)\mid \bar
J_{\alpha\beta}(0)\mid
 0\rangle}{m_{K_2^*(f_2)(a_2)}^2-q^2}
&+& \cdots,
\end{eqnarray}
where dots indicate the contributions of the higher and continuum
states. The matrix element $\langle 0 \mid J_{\mu\nu}(0)\mid
K_2^*(f_2)(a_2)\rangle$ creating the tensor mesons from  vacuum
can be written in terms of the decay constant,
$f_{K_2^*(f_2)(a_2)}$ as
\begin{eqnarray}\label{matrixelement}
\langle 0 \mid J_{\mu\nu}(0)\mid
K_2^*(f_2)(a_2)\rangle=f_{K_2^*(f_2)(a_2)}
m_{K_2^*(f_2)(a_2)}^3\varepsilon^{(\lambda)}_{\mu\nu},
\end{eqnarray}
where $\varepsilon^{(\lambda)}_{\mu\nu}$ is the polarization
tensor. Using Eq. (\ref{matrixelement}) in Eq. (\ref{phen1}), the
final expression of the physical side is obtained as
\begin{eqnarray}\label{phen2}
\Pi _{\mu\nu,\alpha\beta}(q,T)=\frac{f_{K_2^*(f_2)(a_2)}^2
m_{K_2^*(f_2)(a_2)}^6} {m_{K_2^*(f_2)(a_2)}^2-q^2}
\left\{\frac{1}{2}(g_{\mu\alpha}g_{\nu\beta}+g_{\mu\beta}g_{\nu\alpha})\right\}+
\mbox{other structures}+...,
\end{eqnarray}
where the only structure that we will use in our calculations
has been shown explicitly. To obtain the above expression we have used the
summation over polarization tensors as
\begin{eqnarray}\label{polarizationt1}
\sum_{\lambda}\varepsilon_{\mu\nu}^{(\lambda)}\varepsilon_{\alpha\beta}^{*(\lambda)}=\frac{1}{2}T_{\mu\alpha}T_{\nu\beta}+
\frac{1}{2} T_{\mu\beta}T_{\nu\alpha}
-\frac{1}{3}T_{\mu\nu}T_{\alpha\beta},
\end{eqnarray}
where
\begin{eqnarray}\label{polarizationt2}
T_{\mu\nu}=-g_{\mu\nu}+\frac{q_\mu q_\nu}{m_{K_2^*(f_2)(a_2)}^2}.
\end{eqnarray}

Now we concentrate on the OPE side of the thermal correlation function. In OPE representation,
the coefficient of the selected structure can be separated into  perturbative and non-perturbative parts
\begin{eqnarray}\label{piqcd}
\Pi(q,T)=\Pi^{pert}(q,T)+\Pi^{non-pert}(q,T).
\end{eqnarray}
The perturbative or short-distance  contributions are calculated using the
perturbation theory. This part in spectral representation is written as 
\begin{eqnarray}\label{QCD Side}
\Pi^{pert}(q,T) =\int ds\frac{\rho(s)}{s-q^2},
\end{eqnarray}
where $\rho(s)$ is the spectral density and it is given by the
imaginary part of the correlation function, i.e.,
\begin{eqnarray}\label{QCD Side}
\rho(s)=\frac{1}{\pi}Im[\Pi^{pert}(s,T)].
\end{eqnarray}
The non-perturbative or long-distance  contributions are
represented in terms of  thermal expectation values of the
quark and gluon condensates as well as thermal average of the
energy density. Our main task in the following is to calculate the spectral density as well as the non-perturbative contributions.
 For this aim we use
the explicit forms of the  interpolating currents for the tensor mesons  in Eq.
(\ref{correl.func.101}). After 
contracting out all quark fields using the Wick's theorem, we get
\begin{eqnarray}\label{qcdsidek2s}
\Pi _{\mu\nu,\alpha\beta}^{K_2^{*}}(q,T)&=&-\frac{i}{4}\int
d^{4}xe^{iq\cdot(x-y)} \Bigg\{Tr\Big[S_s(y-x)\gamma_\mu\olra{\cal
D}_{\nu}(x)\olra{\cal D}_{\beta}(y)S_d(x-y)\gamma_\alpha\Big]
\nonumber\\&+&\left[\beta\leftrightarrow\alpha\right]+\left[\nu\leftrightarrow\mu\right]
+\left[\beta\leftrightarrow\alpha,
\nu\leftrightarrow\mu\right]\Bigg\},
\end{eqnarray}
and
\begin{eqnarray}\label{qcdsidef2a2}
\Pi _{\mu\nu,\alpha\beta}^{f_2(a_2)}(q,T)&=&-\frac{i}{8}\int
d^{4}xe^{iq\cdot(x-y)} \Bigg\{Tr\Big[S_u(y-x)\gamma_\mu\olra{\cal
D}_{\nu}(x)\olra{\cal
D}_{\beta}(y)S_u(x-y)\gamma_\alpha\nonumber\\&+&
S_d(y-x)\gamma_\mu\olra{\cal D}_{\nu}(x)\olra{\cal
D}_{\beta}(y)S_d(x-y)\gamma_\alpha\Big]+\left[\beta\leftrightarrow\alpha\right]
+\left[\nu\leftrightarrow\mu\right] \nonumber\\&+&
\left[\beta\leftrightarrow\alpha,
\nu\leftrightarrow\mu\right]\Bigg\}.
\end{eqnarray}
To proceed we need to know the thermal light quark
propagator $S_{q=u,d,s}(x-y)$ in coordinate space which is given as
\cite{arzu2,ZGWang}:
\begin{eqnarray}\label{lightquarkpropagator}
S_{q}^{ij}(x-y)&=& i\frac{\!\not\!{x}-\!\not\!{y}}{
2\pi^2(x-y)^4}\delta_{ij}
-\frac{m_q}{4\pi^2(x-y)^2}\delta_{ij}-\frac{\langle
\bar{q}q\rangle}{12}\delta_{ij} -\frac{(x-y)^{2}}{192} m_{0}^{2}
\langle
\bar{q}q\rangle\Big[1-i\frac{m_q}{6}(\!\not\!{x}-\!\not\!{y})\Big]\delta_{ij}
\nonumber\\
&+&\frac{i}{3}\Big[(\!\not\!{x}-\!\not\!{y})\Big(\frac{m_q}{16}\langle
\bar{q}q\rangle-\frac{1}{12}\langle u\Theta^{f}u\rangle\Big)
+\frac{1}{3}\Big(u\cdot(x-y)\!\not\!{u}\langle
u\Theta^{f}u\rangle\Big)\Big]\delta_{ij}
\nonumber\\
&-&\frac{ig_s}{32\pi^{2}(x-y)^{2}}
G_{\mu\nu}\Big((\!\not\!{x}-\!\not\!{y})\sigma^{\mu\nu}+\sigma^{\mu\nu}(\!\not\!{x}-\!\not\!{y})\Big)\delta_{ij},
\end{eqnarray}
where $\langle\bar{q}q\rangle$ is the temperature-dependent quark condensate,  $\Theta^{f}_{\mu\nu}$ is the fermionic part of the energy
momentum tensor and  $u_{\mu}$ is the four-velocity of the heat
bath. In the rest frame of the heat bath, $u_{\mu}=(1,0,0,0)$ and
$u^2=1$.

The  next step is to use the expressions of the propagators and
apply the derivatives with respect to $x$ and $y$ in Eqs.
(\ref{qcdsidek2s}) and (\ref{qcdsidef2a2}).  After
lengthy but straightforward calculations  the  spectral densities at different channels are obtained as
\begin{eqnarray}\label{spectraldenstyK2s}
\rho_{K_{2}^*}(s)=N_c\Big(\frac{m_d m_s
s}{32\pi^2}+\frac{s^{2}}{160\pi^{2}}\Big),
\end{eqnarray}
and
\begin{eqnarray}\label{spectraldenstyf2a2}
\rho_{f_2
(a_2)}(s)=N_c\Big(\frac{(m_{u}^{2}+m_{d}^{2})s}{96\pi^2}+\frac{s^{2}}{160\pi^{2}}\Big),
\end{eqnarray}
where $N_c=3$ is the number of colors.  From a similar way, for the non-perturbative
contributions we get
\begin{eqnarray}\label{nonpertK2s}
\Pi^{non-pert}_{K_2^{*}}(q,T)=\frac{(6m_s-5m_d)m_0^{2}}{144q^{2}}\langle\bar{d}d\rangle+\frac{(6m_d-5m_s)m_0^{2}}{144q^{2}}\langle\bar{s}s\rangle
-\frac{2\langle u\Theta^{f}u\rangle(q\cdot u)^{2}}{9q^{2}},
\end{eqnarray}
and
\begin{eqnarray}\label{nonpertf2a2}
\Pi^{non-pert}_{f_2(a_2)}(q,T)=\frac{m_d
m_0^{2}}{144q^{2}}\langle\bar{d}d\rangle+\frac{m_u
m_0^{2}}{144q^{2}}\langle\bar{u}u\rangle -\frac{2\langle
u\Theta^{f}u\rangle(q\cdot u)^{2}}{9q^{2}}.
\end{eqnarray}

After matching the hadronic and OPE representations, applying Borel transformation with respect to $q^2$ and performing continuum subtraction we obtain
the following temperature-dependent sum rule
\begin{eqnarray}\label{rhomatching}
&&f_{K_2^*(f_2)(a_2)}^2(T)m_{K_2^*(f_2)(a_2)}^6(T)exp\Big[\frac{-m_{K_2^*(f_2)(a_2)}^{2}(T)}{M^{2}}\Big]\nonumber\\
 &&=\int_{(m_q+m_d)^2}^{s_0(T)} ds \Big\{\rho_{K_2^*(f_2)(a_2)}(s)exp\Big[\frac{-s}{M^{2}}\Big]\Big\}
+\hat {\cal B}\Pi_{K_2^*(f_2)(a_2)}^{non-pert}(q,T),
\end{eqnarray}
where $\hat {\cal B}$ denotes the Borel transformation with respect to $q^2$, $M^2$ is the Borel mass parameter, $s_0(T)$ is the temperature-dependent continuum
threshold and $m_q$ can be $m_u$, $m_d$ or $m_s$ depending on the kind of tensor meson. The temperature-dependent mass of the considered tensor states is found as
\begin{eqnarray}\label{rhomass}
m_{K_2^*(f_2)(a_2)}^2(T) =\frac{\int_{(m_q+m_d)^2}^{s_0(T)}ds\Big\{
\rho_{K_2^*(f_2)(a_2)}(s)~s~exp\Big[\frac{-s}{M^{2}}\Big]\Big\}-\frac{d}{d(\frac{1}{M^2})}\Big[\hat {\cal B}\Pi_{K_2^*(f_2)(a_2)}^{non-pert}\Big]}
{\int_{(m_q+m_d)^2}^{s_0(T)}ds \Big\{\rho_{K_2^*(f_2)(a_2)}(s)
exp\Big[\frac{-s}{M^{2}}\Big]\Big\}+\hat {\cal B}\Pi_{K_2^*(f_2)(a_2)}^{non-pert}}.
\end{eqnarray}

\section{Numerical Analysis}
In this section, we discuss the sensitivity of the masses and decay
constants of the $f_2$, $a_2$ and $K_2^{*}$ tensor mesons to 
temperature and compare the results obtained at zero temperature with the predictions of vacuum sum rules \cite{Aliev,Aliev2} as well as
the existing experimental data \cite{J. Beringer}. For this aim,
we use some input parameters as:
$m_u=(2.3_{-0.5}^{+0.7}) MeV$, $m_d=(4.8_{-0.3}^{+0.7}) MeV$ and
$m_s=(95\pm5) MeV$ \cite{J. Beringer}, 
$\langle0|\overline{u}u|0\rangle=\langle0|\overline{d}d|0\rangle=-(0.24\pm0.01)^3~$GeV$^3$
\cite{B.L.Ioffe} and
$\langle0|\overline{s}s|0\rangle=0.8\langle0|\overline{u}u|0\rangle$
\cite{S. Narison}.

In further analysis we need to know the expression of the light quark condensate
 at finite temperature calculated at different works (see for instance \cite{barduci,ayala}).  In the present study, we use the parametrization  obtained in \cite{ayala} which is also consistent
with the  lattice results \cite{latt1,latt2}.
%
%
For the  temperature-dependent continuum threshold  we also use the parametrization obtained in \cite{ayala} in terms of the temperature-dependent light-quark condensate and continuum threshold in vacuum ($s_0$).
%
%
 The continuum threshold $s_{0}$ is not
completely arbitrary and is correlated with the energy of the
first excited state with the same quantum numbers as the chosen
interpolating currents.  Our analysis reveals that in the
intervals $(2.2-2.5)~ GeV^2$, $(2.4-2.7) ~GeV^2$ and
$(3.0-3.3) ~GeV^2$ respectively for $f_2$, $a_2$
and $K_2^{*}$ channels the results weakly depend on the continuum threshold. Hence, we consider these intervals as working regions of $s_{0}$ for the channels under consideration.

 According to the general philosophy of the method used the  physical quantities under consideration should also be practically independent of the  Borel mass
parameter $M^2$. The working region for this parameter are determined by requiring that
not only the higher state and continuum contributions are
suppressed, but also the contribution of the highest order operator
are small.  Taking into account  these conditions we find that in the interval
$1.4 ~GeV^2 \leq M^2\leq 3~ GeV^2$ ,
the results weakly depend on $M^2$. Figure 1 indicates the dependence of
 the masses and decay constants on the Borel mass parameter at zero temperature. From this figure we see that the results demonstrate good stability with respect to the 
variations of $M^2$ in its working region.
 \begin{figure}[h!]
\begin{center}
\includegraphics[width=8cm]{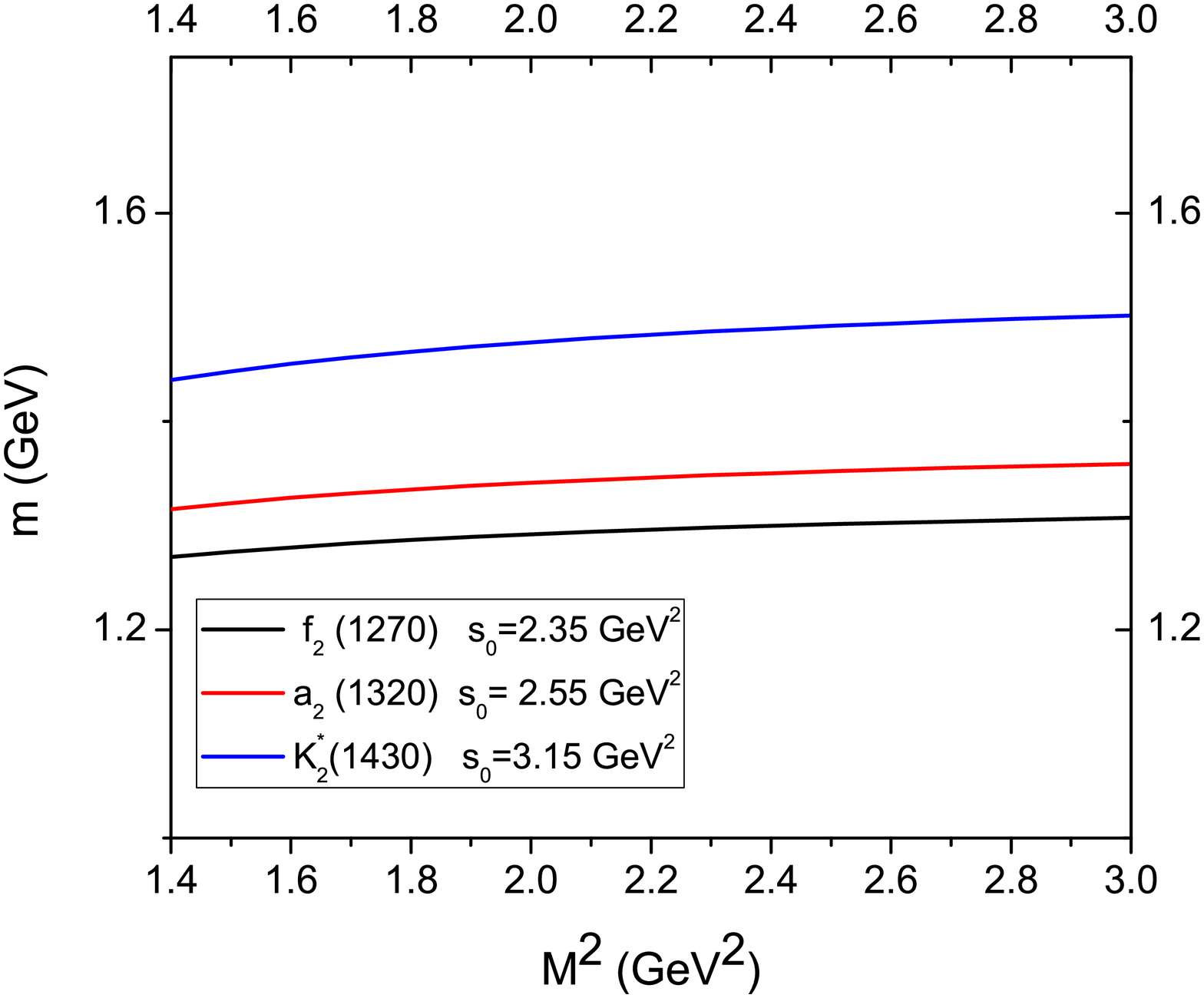}
\includegraphics[width=8cm]{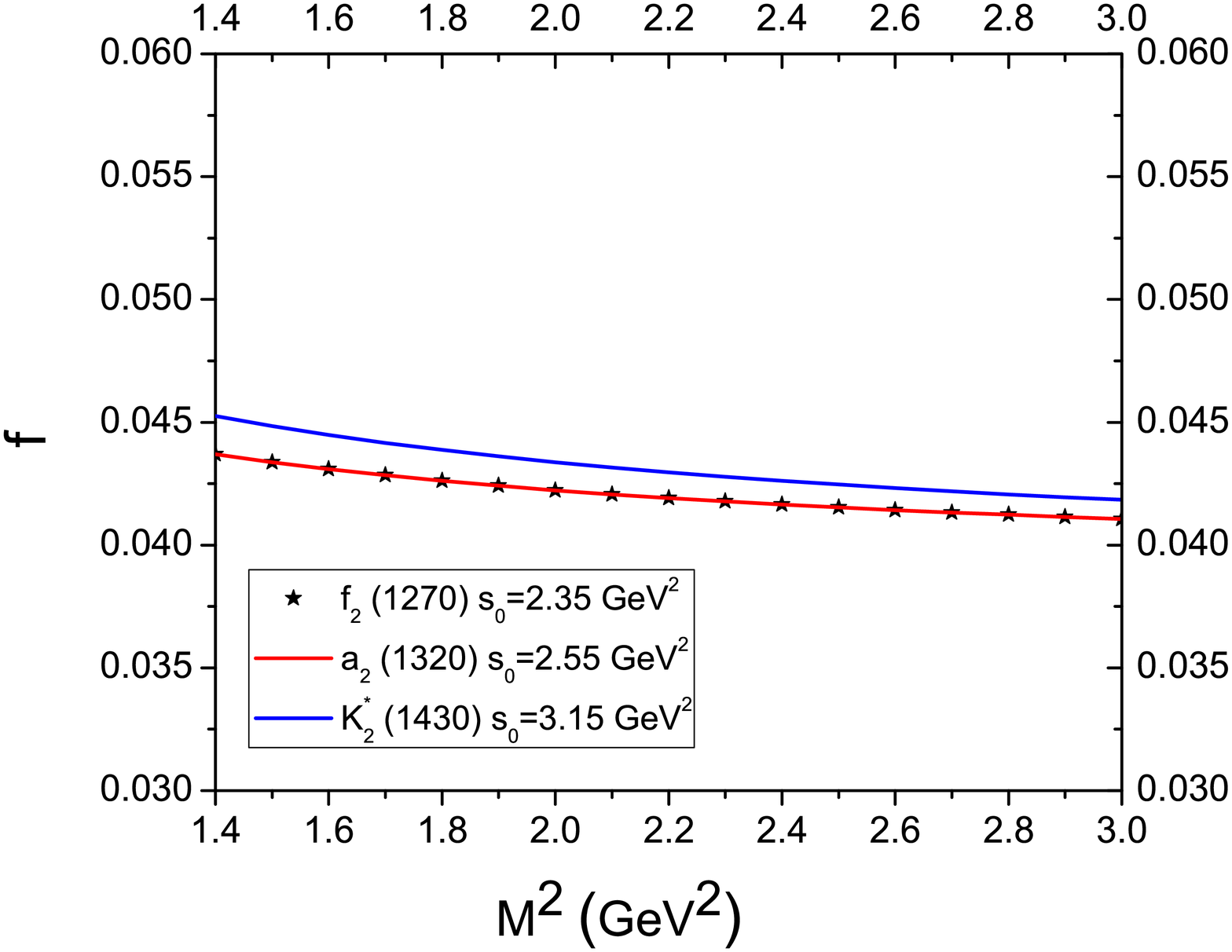}
\end{center}
\caption{Variations of the masses and decay constants of the
$K_2^*(1430)$, $f_2(1270)$ and $a_2(1320)$ mesons with respect to
$M^2$  at fixed values of the continuum threshold and at zero
temperature.} \label{Diagrams1}
\end{figure}
 \begin{figure}[h!]
\begin{center}
\includegraphics[width=8cm]{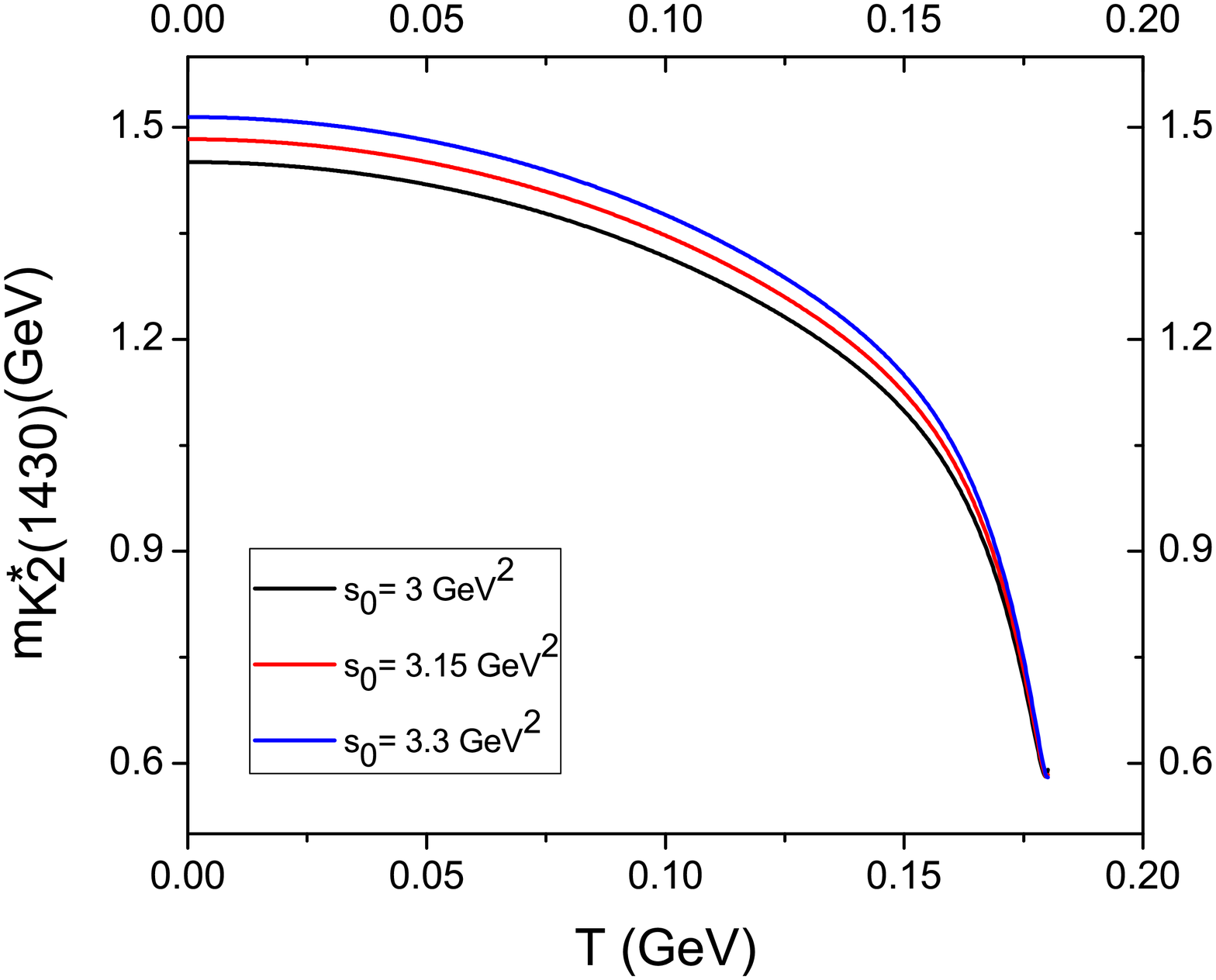}
\includegraphics[width=8cm]{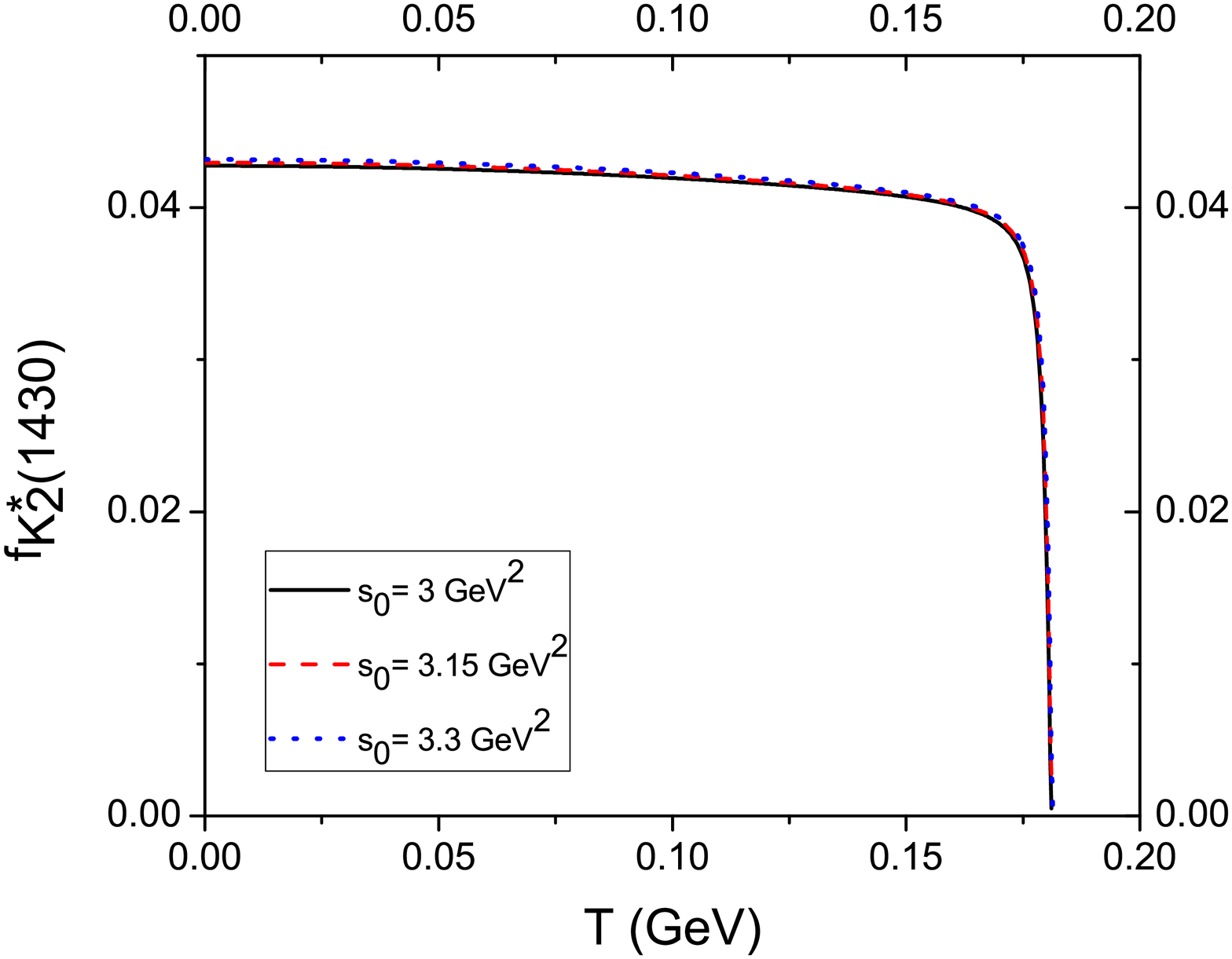}
\end{center}
\caption{Temperature dependence of the mass and decay constant of
the $K_{2}^*(1430)$ meson.
} \label{Diagrams1}
\end{figure}

 \begin{figure}[h!]
\begin{center}
\includegraphics[width=8cm]{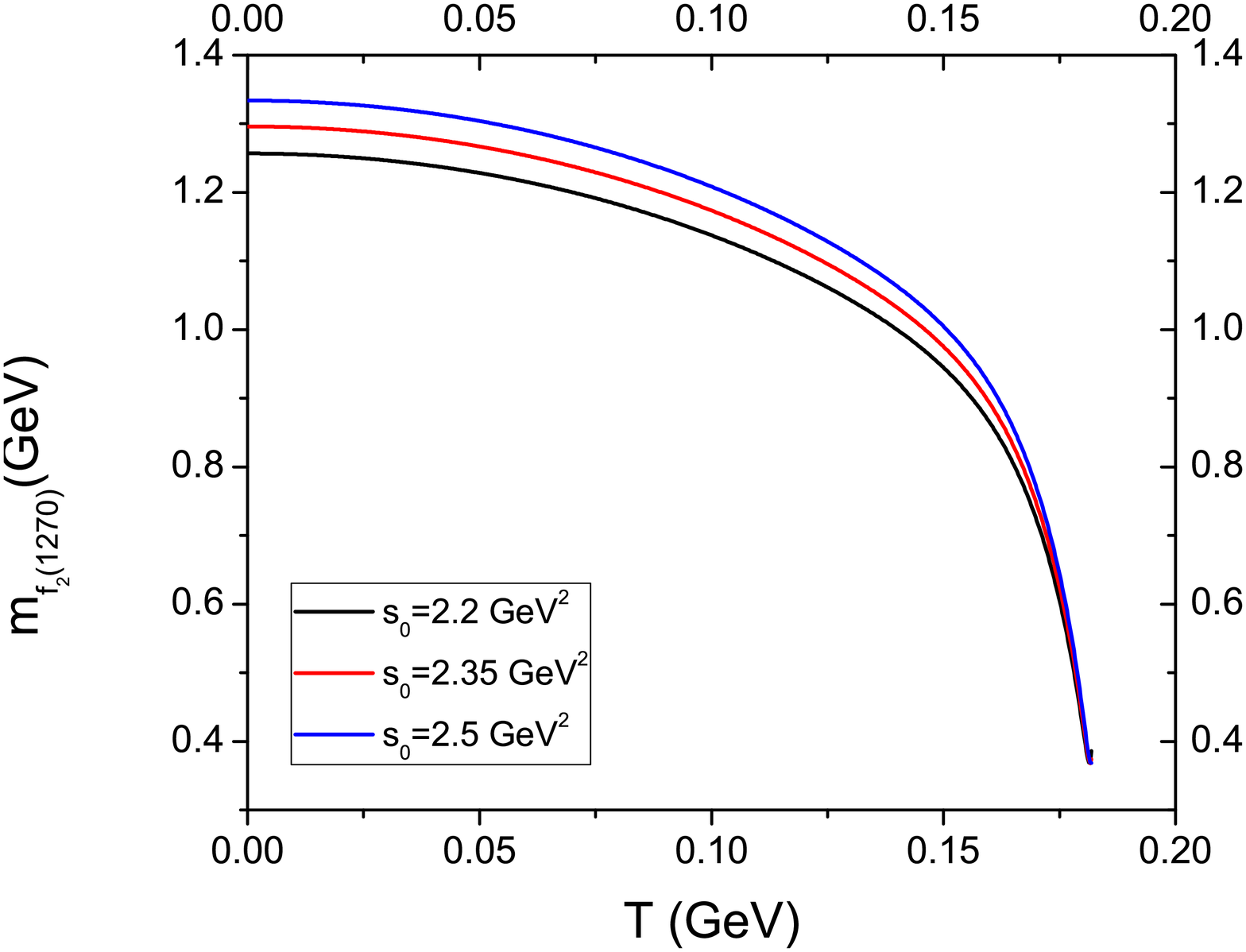}
\includegraphics[width=8cm]{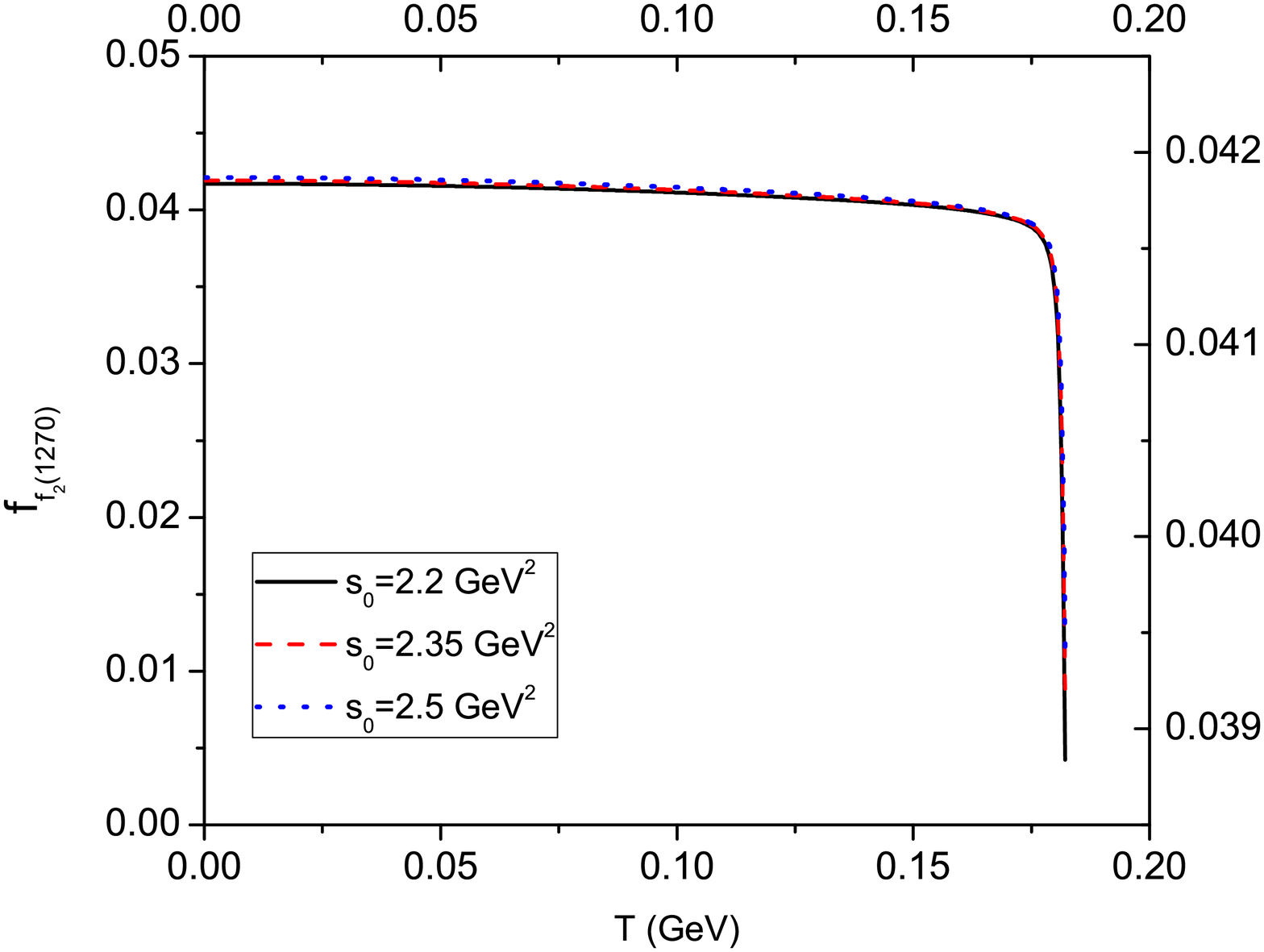}
\end{center}
\caption{Temperature dependence of the mass and decay constant of
the $f_2(1270)$ meson.
} \label{Diagrams1}
\end{figure}

 \begin{figure}[h!]
\begin{center}
\includegraphics[width=8cm]{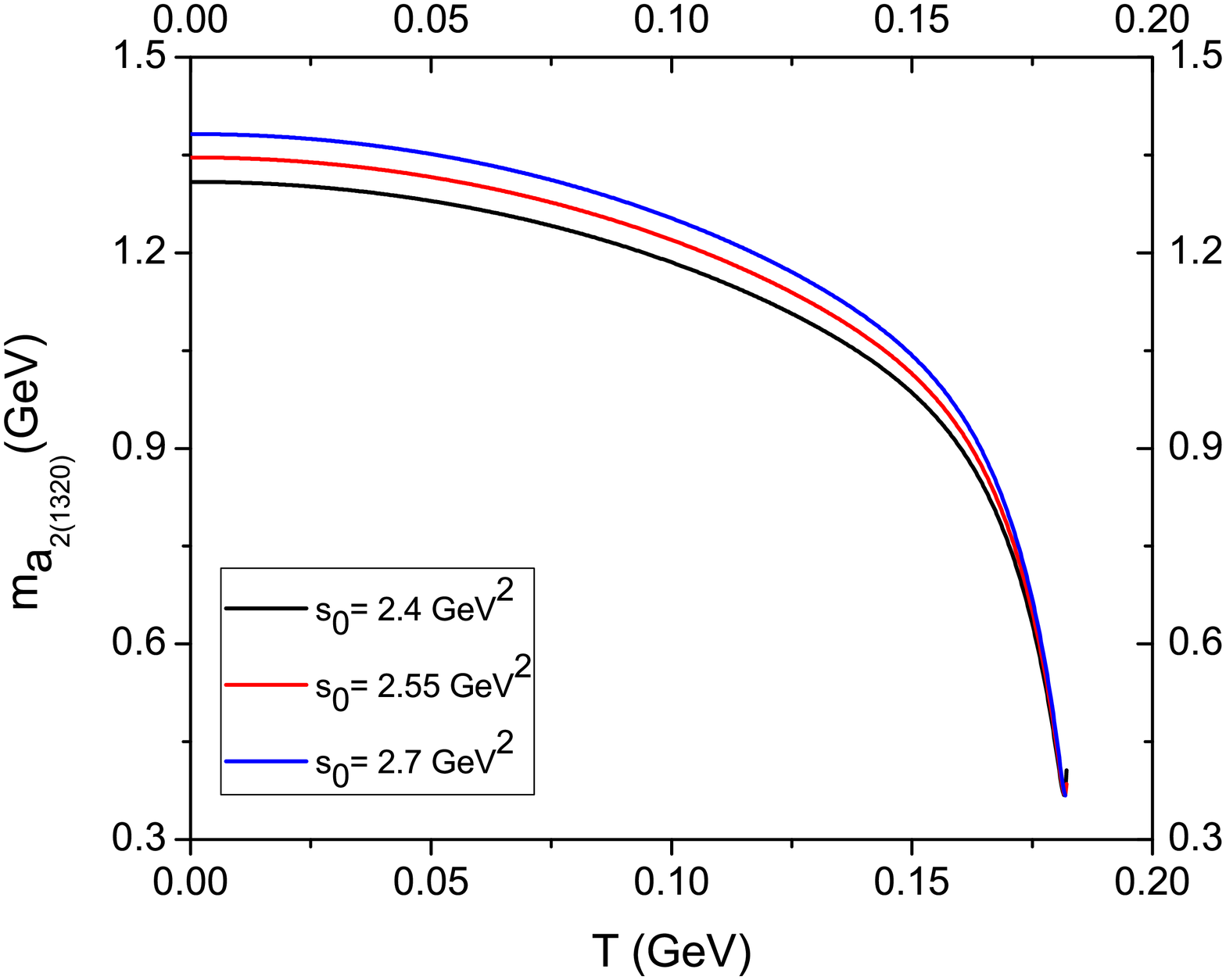}
\includegraphics[width=8cm]{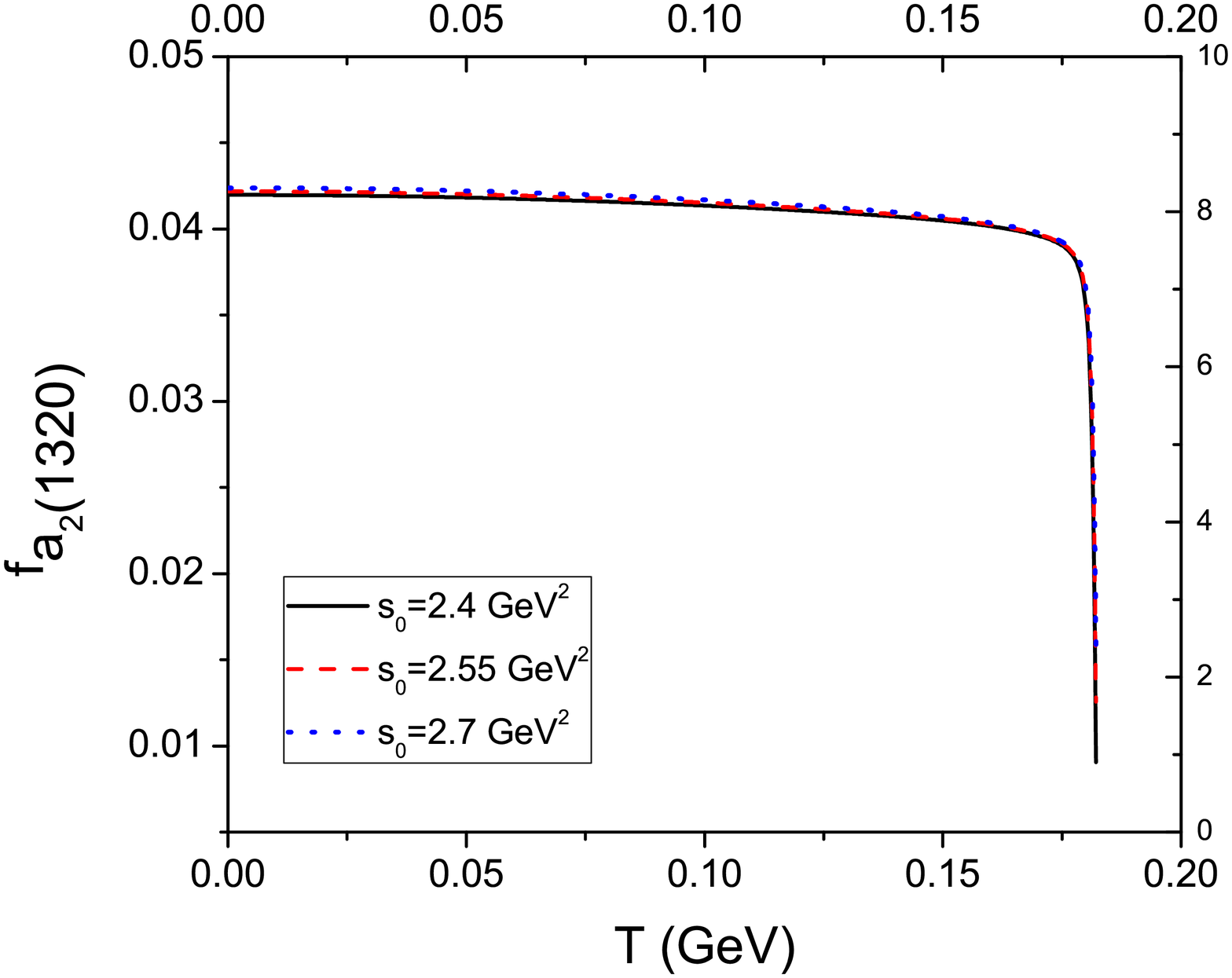}
\end{center}
\caption{Temperature dependence of the mass and decay constant of
the $a_2(1320)$ meson.
} \label{Diagrams1}
\end{figure}

Now, we proceed to discuss how the physical quantities under consideration behave in terms of temperature in the working regions of the auxiliary parameters $M^2$ and $s_0$. For this aim,  we present the dependence of the  masses and
decay constants on temperature at $M^2=2.2~GeV^2$ in figures 2, 3 and 4.
Note that, we plot these figures up to the temperature that the temperature-dependent continuum threshold vanishes, i.e., $T\simeq183~ MeV$.  From these figures, we see that the masses and decay constants  diminish by increasing the temperature.
 Near to the   temperature $T\simeq183~ MeV$, the decay
constants of the $f_{2}(1270)$, $a_{2}(1320)$ and $K_2^*(1430)$   decrease with amount of $81\%$, $70\%$ and $85\%$    compared to their vacuum values, respectively. While, the
masses decrease about $70\%$, $72\%$ and $60\%$ for
$f_{2}(1270)$, $a_{2}(1320)$ and $K_2^*(1430)$ states,
respectively.
 
\begin{table}[h]
\renewcommand{\arraystretch}{1.5}
\addtolength{\arraycolsep}{3pt}
$$
\begin{array}{|c|c|c|c|c|c|c|c|}
\hline \hline
         &\mbox{Present\,\,\,Work} & \mbox{Experiment}\,\,& \mbox{Vacuum\,\,Sum\,\,Rules}\,\,\cite{Aliev,Aliev2}\,\, \\
        & & \cite{J. Beringer}&  \mbox{Relativistic\,\, Quark \,\,Model}\,\,\cite{Ebert}\,\, \\
\hline
  \mbox{$m_{K_2^*(1430)}$(GeV)}        &  1.48\pm0.12    &  1.4256\pm0.0015   &1.44\pm0.10~~~\cite{Aliev2},~~~1.424~~~\cite{Ebert}\\
\hline
  \mbox{ $f_{K_2^*(1430)}$}        &  0.043\pm0.002 &  -  & 0.050\pm0.002 ~~\cite{Aliev2}\\
\hline
  \mbox{$m_{f_{2}(1270)}$(GeV)}        &  1.30\pm0.08    &  1.2751\pm0.0012 &  1.25~~~~~\cite{Aliev}\\
\hline
  \mbox{ $f_{f_{2}(1270)}$}        &  0.042\pm0.002 &  -  &  0.040  ~~~~~\cite{Aliev} \\
  \hline
  \mbox{$m_{a_{2}(1320)}$(GeV)}        &  1.35\pm0.11    &  1.3183\pm0.0006&  1.25~~~~~\cite{Aliev}\\
\hline
  \mbox{ $f_{a_{2}(1320)}$}        &  0.042\pm0.002 &  -  &  - \\

                    \hline \hline
\end{array}
$$
\caption{Values of the masses and decay constants of the
 $K_2^*$, $f_2$ and $a_{2}$ mesons at zero temperature.} \label{tab:lepdecconst}
\renewcommand{\arraystretch}{1}
\addtolength{\arraycolsep}{-1.0pt}
\end{table}

Our final task is to compare the  results of this work obtained at zero temperature with those of the  vacuum sum rules as
well as other existing theoretical predictions and experimental data. This comparison is made
in table 1. From this table we see that the results on the masses and decay constants obtained at zero temperature
 are roughly consistent with existing experimental data as
well as the vacuum sum rules and relativistic quark model predictions  
within the uncertainties. Our predictions on the decay constants
of the light tensor mesons can be checked in future experiments. The results obtained in the present work can be used in theoretical determination of the electromagnetic properties of the light  tensor mesons
as well as their weak decay parameters and their strong couplings with other hadrons. Our results on the thermal behavior of the masses and decay constants can also be useful in analysis of the results of future heavy ion collision experiments.

\section{Acknowledgment}
This work has been supported in part by the Scientific and Technological
Research Council of Turkey (TUBITAK) under the research projects
 110T284 and 114F018.

\section{Conflict of Interests}
The authors declare that there is no conflict
of interests regarding the publication of this article.


\begin{thebibliography}{99}

%
%
%
%
%
%
%
%
%
%
%
%
%
%
%
%
\bibitem{G.Brown} G. E. Brown and M. Rho, Phys. Rep. \textit{269}, 333 (1996).
%

%
\bibitem{Shifman1} M. A. Shifman, A. I. Vainshtein,  V. I. Zakharov,  Nucl. Phys. \textit{B147}, 385 (1979).
 \bibitem{Shifman2} M. A. Shifman, A. I. Vainshtein,  V. I. Zakharov, Nucl. Phys. \textit{B147}, 448 (1979).
%
\bibitem{Bochkarev} A. I. Bochkarev,  M. E. Shaposhnikov,  Nucl. Phys.
\textit{B268}, 220 (1986).
%
%
\bibitem{Hatsuda} T. Hatsuda, Y. Koike and S. H. Lee, Nucl. Phys. \textit{B394}, 221 (1993).
%
\bibitem{Mallik2} S. Mallik, Phys. Lett. \textit{B416}, 373 (1998).
%
\bibitem{Shuryak} E.V. Shuryak, Rev. Mod. Phys. \textit{65}, 1 (1993).

\bibitem{Mallik} S. Mallik and S. Sarkar, Eur. Phys. J. \textit{C25}, 445 (2002).
%

\bibitem{Waas} T. Waas,  N. Kaiser, W. Weise,  Phys. Lett. \textit{B365}, 12 (1996).
%
\bibitem{Tolos} L. Tolos, D. Cabrera and A. Ramos, Phys. Rev. \textit{C78}, 045205
(2008).
%
%

\bibitem{Veliev} E. V. Veliev, J. Phys. \textit{G35}, 035004
(2008).
%
%
\bibitem{Veliev3} E. V. Veliev, K. Azizi, H. Sundu, N. Aksit, J. Phys. \textit{G39}, 015002 (2012).
\bibitem{Velievuc} E. V. Veliev, G. Kaya, Eur. Phys. J. \textit{C63}, 87 (2009).
%
\bibitem{Klingl} F. Klingl, S. Kim, S. H. Lee, P. Morath
and W. Weise, Phys. Rev. Lett. \textit{82}, 3396 (1999).
%
\bibitem{Loewe} C. A. Dominguez, M. Loewe,
J.C. Rojas, JHEP \textit{08}, 040 (2007).
 \bibitem{Loewe1} C. A. Dominguez, M. Loewe, Phys. Lett. \textit{B233}, 201 (1989).
%
%
\bibitem{arzu} E. V. Veliev, K. Azizi, H. Sundu, G. Kaya, A. T\"{u}rkan, Eur. Phys. J. \textit{A47}, 110 (2011).
%
\bibitem{arzu2} K. Azizi, H. Sundu, A. T\"{u}rkan and E. V. Veliev, J. Phys. \textit{G41}, 035003 (2014).
\bibitem{Aliev} T. M. Aliev, M. A. Shifman, Phys. Lett. \textit{B112}, 401 (1982).
\bibitem{Ebert} D. Ebert, R. N. Faustov and V. O. Galkin, Phys. Rev. \textit{D79}, 114029 (2009).
\bibitem{Aliev2} T. M. Aliev, K. Azizi and V. Bashiry, J. Phys. \textit{G37}, 025001 (2010).
%
%
\bibitem{ZGWang}Z. G. Wang, Z. C. Liu, X. H. Zhang,
Eur. Phys. J. \textit{C64},  373 (2009).
%
\bibitem{J. Beringer} J. Beringer et al. (Particle Data Group), Phys. Rev. \textit{D86}, 010001 (2012).
%
\bibitem{B.L.Ioffe} B. L. Ioffe, Prog. Part. Nucl. Phys. \textit{56}, 232 (2006).
%
\bibitem{S. Narison} S. Narison, Phys. Lett. \textit{B605}, 319 (2005).
%
%
\bibitem{barduci} A.  Barducci, R. Casalbuoni, S. De Curtis, R. Gatto, G. Pettini, Phys. Rev. \textit{D46},  2203 (1992). 
\bibitem{ayala} A. Ayala, A. Bashir, C. A. Dominguez, E. Gutierrez, M. Loewe, A. Raya, Phys. Rev. \textit{D84}, 056004 (2011).
\bibitem{latt1} A. Bazavov et al., Phys. Rev. \textit{D80}, 014504 (2009).
\bibitem{latt2} M. Cheng et al., Phys. Rev. \textit{D81}, 054504 (2010). 

\end{thebibliography}
\end{document}